# Trajectory-based interpretation of laser light diffraction by a sharp edge

Milena D. Davidović[1], Miloš D. Davidović[2], Ángel S. Sanz[3], Mirjana Božić[4] and Darko Vasiljević[4]

[1]Faculty of Civil Engineering, University of Belgrade, Serbia
[2]Vinča Institute of Nuclear Sciences, University of Belgrade, Serbia
[3]Department of Optics, Universidad Complutense de Madrid, Spain
[4]Institute of Physics, University of Belgrade, Serbia

**Abstract**

In the diffraction pattern produced by a half-plane sharp edge when it obstructs the passage of a laser beam, two characteristic regions are noticeable. There is a central region, where it can be noticed the diffraction of laser light in the region of geometric shadow, while intensity oscillations are observed in the non-obstructed area. On both sides of the edge, there are also very long light traces along the normal to the edge of the obstacle. The theoretical explanation to this phenomenon is based on the Fresnel-Kirchhoff diffraction theory applied to the Gaussian beam propagation behind the obstacle. Here we have supplemented this explanation by considering electromagnetic flow lines, which provide a more complete interpretation of the phenomenon in terms of electric and magnetic fields and flux lines, and that can be related, at the same time, with average photon paths.

> Light displays a characteristic dual nature**.** **I**t appears both as a wave motion but also as a stream of discrete particles of light, as photons. When a photon hits a material, it can emit one and only one electron. Light may be described by classical optics, but observing it is always based on the absorption of one quantum of energy.
>
> *Stig Stenholm in the Presentation Speech for the 2005 Nobel Prize in Physics awarded to Roy Glauber, John Hall and Theodor Hänsch*

## 1. Introduction

Bohmian mechanics enables the visualization and interpretation of the quantum mechanical behavior of particles with a mass through trajectories associated with the probability current density [1]. Analogously, the electromagnetic field also admits a hydrodynamic formulation when the existence of a suitably defined photon wave function is assumed [2]. This formulation gives the possibility to interpret the optical phenomena in a picturesque way through "photon trajectories", which describe the evolution of the electromagnetic energy density, particularly behind obstacles, where it may result of much interest for the information on diffraction and interference such trajectories provide. This approach, based on trajectories, has been used in the analysis of Young's double slit diffraction [3-5], the Poisson-Arago spot [5,6], the Arago-Fresnel laws [7], the modes and energy propagation in optical and microwave waveguides [8,9] or, more recently, the high-power ultrashort laser-pulse propagation in nonlinear dissipative media [10].

A group of scientists from the University of Toronto, under the guidance of professor Steinberg, has been able to experimentally determine the mean paths of single photons in Young's experiment [11]. The trajectories determined in this way were in agreement with trajectories theoretically previously anticipated [2,3] and evaluated [5]. The achievement of Steinberg's group was selected by Physics World as the top breakthrough in physics for the year 2011, as a discovery that is shifting the moral of quantum measurement [12].

In this work, we tackle the issue of diffraction by a half-plane sharp edge. The theoretical solution for the diffraction of a plane wave by the edge of a perfectly conducting plane was given by Sommerfeld in 1896 [13], which has become a standard starting point in solving diffraction problems for various two dimensional obstacles [14,15]. The diffraction of a Gaussian beam by an edge has been studied since the 1960s [16], although attention was mainly focused on the bright central part [17-19] of the diffraction image, while the less pronounced long side tails have not been analyzed until more recently [20,21]. This less pronounced tail extends with decreasing intensity along the diffraction plane on both sides of the diffracted-beam axis. Here we use the photon trajectory approach to interpret the diffraction pattern observed on a screen allocated beyond the place where the sharp edge produces the diffraction of an incident Gaussian laser beam, such as a razor blade.

This work is organized as follows. The theoretical approach of the electromagnetic energy flow lines (photon trajectories) is presented in Sec. 2. To be self-contained, first a brief overview of the photon trajectory approach is presented, and then the particular application to the problem of the diffraction by a half-plane edge is introduced. The description of the edge-diffraction experiment performed in the laboratory is given in Sec. 3, showing some of the outcomes obtained. Specifically, here we have used a razor blade to produce in the laboratory the edge diffraction patterns that are later on analyzed by means of photon trajectories. Based on the experimental outcomes reported in Sec. 3, and making use of the theory previously introduced in Sec. 2, in Sec. 4 we present and discuss our main results. These results show a good agreement between the theoretical predictions of the diffraction theory and the space evolution of swarms of electromagnetic flow lines, particularly when their arrivals are collected and represented in the form of a histogram. Finally, a concluding summary is given in Sec. 5.

## 2. Electromagnetic energy flow approach

### 2.1. Field and trajectory equations

Electric and magnetic fields in vacuum obey the following wave equations:

$$\nabla^2 \vec{E}(\vec{r},t) - \frac{1}{c^2}\frac{\partial^2 \vec{E}(\vec{r},t)}{\partial t^2} = 0, \qquad (1)$$

$$\nabla^2 \vec{H}(\vec{r},t) - \frac{1}{c^2}\frac{\partial^2 \vec{H}(\vec{r},t)}{\partial t^2} = 0, \qquad (2)$$

where c is the speed of light in vacuum, $\vec{r}$ is the position vector and $t$ is time. For a monochromatic electromagnetic (EM) wave the electric and magnetic fields are given by $\vec{E}(\vec{r},t) = \vec{E}(\vec{r})e^{-i\omega t}$ and $\vec{H}(\vec{r},t) = \vec{H}(\vec{r})e^{-i\omega t}$, respectively. So, from Eqs. (1) and (2), it follows that the complex amplitudes $\vec{E}$ and $\vec{H}$ satisfy the Helmholtz equations

$$\nabla^2 \vec{E}(\vec{r}) + k^2 \vec{E}(\vec{r}) = 0, \qquad (3)$$

$$\nabla^2 \vec{H}(\vec{r}) + k^2 \vec{H}(\vec{r}) = 0, \qquad (4)$$

where $k = \omega/c = 2\pi/\lambda$. The electromagnetic energy (EME) flow lines, describing the flow or propagation of the corresponding EM energy in vacuum, are determined using the energy flux vector, given by the real part of the (time-averaged) complex Poynting vector,

$$\vec{S}(\vec{r}) = \frac{1}{2} Re[\vec{E}(\vec{r}) \times \vec{H}^*(\vec{r})], \qquad (5)$$

from the equation

$$\frac{d\vec{r}}{ds} = \frac{\vec{S}(\vec{r})}{S(\vec{r})} = \frac{\vec{S}(\vec{r})}{cU(\vec{r})}, \qquad (6)$$

where ds is an elementary arc length along the EME flow line, and $U$ is the time-averaged energy density

$$U(\vec{r}) = \frac{1}{4}[\epsilon_0 \vec{E}(\vec{r})\vec{E}^*(\vec{r}) + \mu_0 \vec{H}(\vec{r})\vec{H}^*(\vec{r})]. \qquad (7)$$

### 2.2 Diffraction by a half-plane edge

The theoretical solution of the diffraction problem is obtained by solving the Helmholtz equation behind the obstacle, so that the boundary conditions at the obstacle are satisfied. The solution can be written in the form of the Fresnel-Kirchhoff integral [15]. Let us consider the incident beam travelling along the y axis, coming to the opaque obstacle located at xOz plane, with the edge along the z axis. For simplicity, we will assume that the incident wave does not depend on the z-coordinate, and it is Gaussian along the x axis. In that case, the Fresnel-Kirchhoff integral reads as

$$\Psi(x,y) = \sqrt{\frac{k}{2\pi y}} e^{-\frac{i\pi}{4}} e^{iky} \int_{-\infty}^{+\infty} \Psi_0(x', 0^+) e^{ik(x-x')^2/2y} dx', \qquad (8)$$

where

$$\Psi_0(x', 0^+) = e^{-\frac{x'^2}{4\sigma^2}}, \tag{9}$$

for $-\infty < x' < +\infty$, in the case of free propagation, and

$$\Psi_0(x', 0^+) = \begin{cases} 0, & x' > 0 \\ e^{-\frac{x'^2}{4\sigma^2}}, & x' \leq 0 \end{cases}. \tag{10}$$

if the laser beam meets the half plane.

After substitution of the ansatz (9) into the functional form (8) we obtain

$$\Psi(x, y) = \sqrt{\frac{k}{2\pi y}} e^{-\frac{i\pi}{4}} e^{ik(y+\frac{x^2}{2y})} e^{\frac{b^2}{4a}} \sqrt{\frac{\pi}{a}}, \tag{11}$$

for the propagation of the free Gaussian, with

$$a = \frac{1}{4\sigma^2} - \frac{ik}{2y}, \quad b = -\frac{ikx}{y}. \tag{12}$$

On the other hand, when the ansatz (10) is substituted into (8), we obtain

$$\Psi(x, y) = \sqrt{\frac{k}{2\pi y}} e^{-\frac{i\pi}{4}} e^{iky} \int_{-\infty}^{0} e^{-\frac{x'^2}{4\sigma^2}} e^{ik(x-x')^2/2y} dx' \tag{13}$$

for propagation behind a half plane. In deriving the solution of the Helmholtz equation in the form (8), it is assumed [15] that $\lambda \ll \sigma$ and $x \ll y$ (paraxial approximation).

The incident wave can be recast, in general, in terms of two components [15]: *H*-polarized, with the magnetic field polarized along the z-axis, and E- polarized, with the electric field polarized along the z-axis, with a phase shift $\phi$ between them. Behind the obstacle the magnetic field of the *H*-polarized wave and the electric field of the *E*-polarized wave are proportional to $\Psi(x,y)$, and read as $\vec{H}_h = A\Psi\vec{e}_z$ and $\vec{E}_e = \frac{Be^{i\phi}}{\epsilon_0 c}\Psi\vec{e}_z$, respectively.

From Maxwell's equations it follows [7] that the electric field of the *H*-polarized wave is given by $\vec{E}_h = \frac{i}{\omega\epsilon_0}\nabla \times \vec{H}_h$, while the magnetic field of the *E*-polarized wave is $\vec{H}_e = -\frac{i}{\omega\mu_0}\nabla \times \vec{E}_e$, so that the total solution behind the obstacles is given by [2,3,7]:

$$\vec{H} = -ik^{-1}Be^{i\phi}\frac{\partial \Psi}{\partial y}\vec{e}_x + ik^{-1}Be^{i\phi}\frac{\partial \Psi}{\partial x}\vec{e}_y + A\Psi\vec{e}_z, \tag{14}$$

$$\vec{E} = \frac{iA}{\epsilon_0\omega}\frac{\partial \Psi}{\partial y}\vec{e}_x - \frac{iA}{\epsilon_0\omega}\frac{\partial \Psi}{\partial x}\vec{e}_y + \frac{kBe^{i\phi}}{\epsilon_0\omega}\Psi\vec{e}_z. \tag{15}$$

The components of the Poynting vector can be expressed as

$$S_x = \frac{i}{4\epsilon_0\omega}(A^2 + B^2)\left(\Psi\frac{\partial \Psi^*}{\partial x} - \Psi^*\frac{\partial \Psi}{\partial x}\right), \tag{16}$$

$$S_y = \frac{i}{4\epsilon_0 \omega}(A^2 + B^2)\left(\Psi \frac{\partial \Psi^*}{\partial y} - \Psi^* \frac{\partial \Psi}{\partial y}\right), \qquad (17)$$

$$S_z = \frac{i}{2\epsilon_0 \omega k} AB \sin\phi \left(\frac{\partial \Psi}{\partial x}\frac{\partial \Psi^*}{\partial y} - \frac{\partial \Psi^*}{\partial x}\frac{\partial \Psi}{\partial y}\right). \qquad (18)$$

In the case we are dealing with here, the solution (8) of the Helmholtz equation satisfies the approximate relations

$$\left|\frac{\partial \Psi}{\partial x}\right| \ll \left|\frac{\partial \Psi}{\partial y}\right|, \quad \left|\frac{\partial \Psi}{\partial y}\right| \approx ik\Psi, \qquad (19)$$

and therefore the EME density (7) becomes proportional to $|\Psi|^2$, since

$$U(\vec{r}) = \frac{\mu_0}{2}(A^2 + B^2)|\Psi|^2. \qquad (20)$$

The photon trajectories are obtained by numerical integration of the differential equations

$$\frac{dx}{dy} = \frac{S_x}{S_y}, \qquad (21)$$

$$\frac{dz}{dy} = \frac{S_z}{S_y}. \qquad (22)$$

In the case of an *E*-polarized or *H*-polarized incident beam, photon paths are located on xOy plane. The initial *x*-coordinates of the flow lines are chosen to be

$$x_i = \sigma F^{-1}(u), \qquad (23)$$

where $F^{-1}(u)$ is the inverse of the Gaussian cumulative distribution function. If the variable $u$ has a uniform distribution on the interval $(0,1)$, $x_i$ will have a Gaussian distribution with variance $\sigma$.

## 3. Experimental setup and diffraction picture

As an illustration of our theoretical approach, we have chosen an experimental setup that can be easily performed at the classroom. As pointed out by Aviani and Erjavec [21], this demonstration is suitable for applying a predict-observe-explain sequence in teaching optics. Specifically, the experimental setup, which is shown in Fig. 1(a), consists of an optical bench containing two supports that hold a green laser pointer with wavelength $\lambda = 532$ nm, an opaque barrier with a razor glued along its vertical edge and an observation screen, which can be accommodated at different distances from the diffracting edge.

As it can be noticed, if the laser beam is left to freely propagate (i.e., the edge is not present in the setup), a circular bright spot is observed on the screen, as shown in Fig. 1(b). However, when the edge is accommodated in the optical bench in such a way that half of the beam is blocked in its way to the observation screen, it can be seen a characteristic diffraction pattern formed by a central highly intense spot surrounded by two long horizontal tails (perpendicular to the edge). This can be seen in panels (c) and (d), which correspond to two different positions of the observation screen with respect to the edge, y = 0.6 m and y = 3 m, respectively.

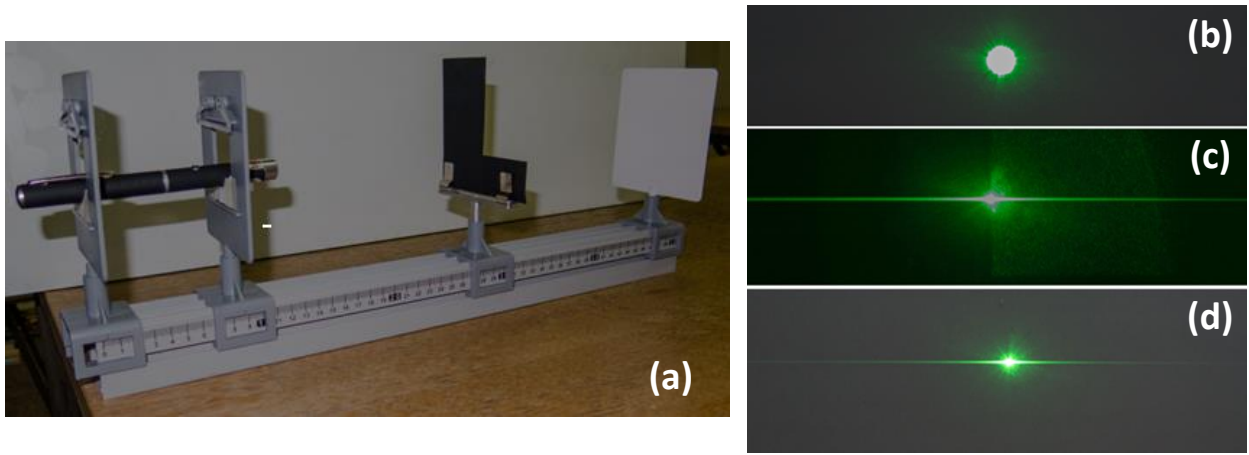

**Figure 1.** (a) Experimental setup to observe diffraction by a straight edge of the light emitted by a green laser pointer ($\lambda = 532$ nm). (b) A green circular spot is observed when the laser beam freely propagates. (c) Diffraction pattern when the edge is introduced in the setup, covering nearly half the incident laser beam, for a distance $y = 0.6$ m between the edge and the observation screen. (d) As in panel (c), but for $y = 3$ m.

## 4. Intensity and EM flow lines for both a free propagating laser beam and an obstructed one by a half plane

Taking into account the wavelength of our laser beam, the intensity curves $I = U(\vec{r})c$ of the freely propagating and obstructed laser beam produced by this theoretical approach can be seen in Fig. 2 for different values of the distance between the edge and the observation screen. As can be seen in Fig. 2(a), the intensity curve of the obstructed beam shows an oscillatory decrease in the non-obstructed region and a simple decrease in the obstructed region. As can be noticed in Figs. 2(b), as the observation screen is displaced further away from the edge, the oscillations on the left-hand side of the diffraction pattern get weaker. These features are well known [16,20] and are similar to the features seen in the diffraction pattern of a plane wave [13-15]. The phenomenon of long tails in Figs. 2(c) and (d) was not noted and studied until recently [20,21]. Anakhov *et al*. [20] analyzed this phenomenon by writing the expression for the electric field as a sum of two contributions, namely a geometrical contribution and a diffraction one. Here, we have evaluated the total electric and magnetic field and the corresponding energy density for both the freely propagating Gaussian beam and the beam obstructed by the half plane. In our opinion, the slower decrease of obstructed Gaussian [see Figs. 2(c) and (d)] explains the phenomenon of long tails on both sides behind the half plane [see Figs. 1(c) and (d)].

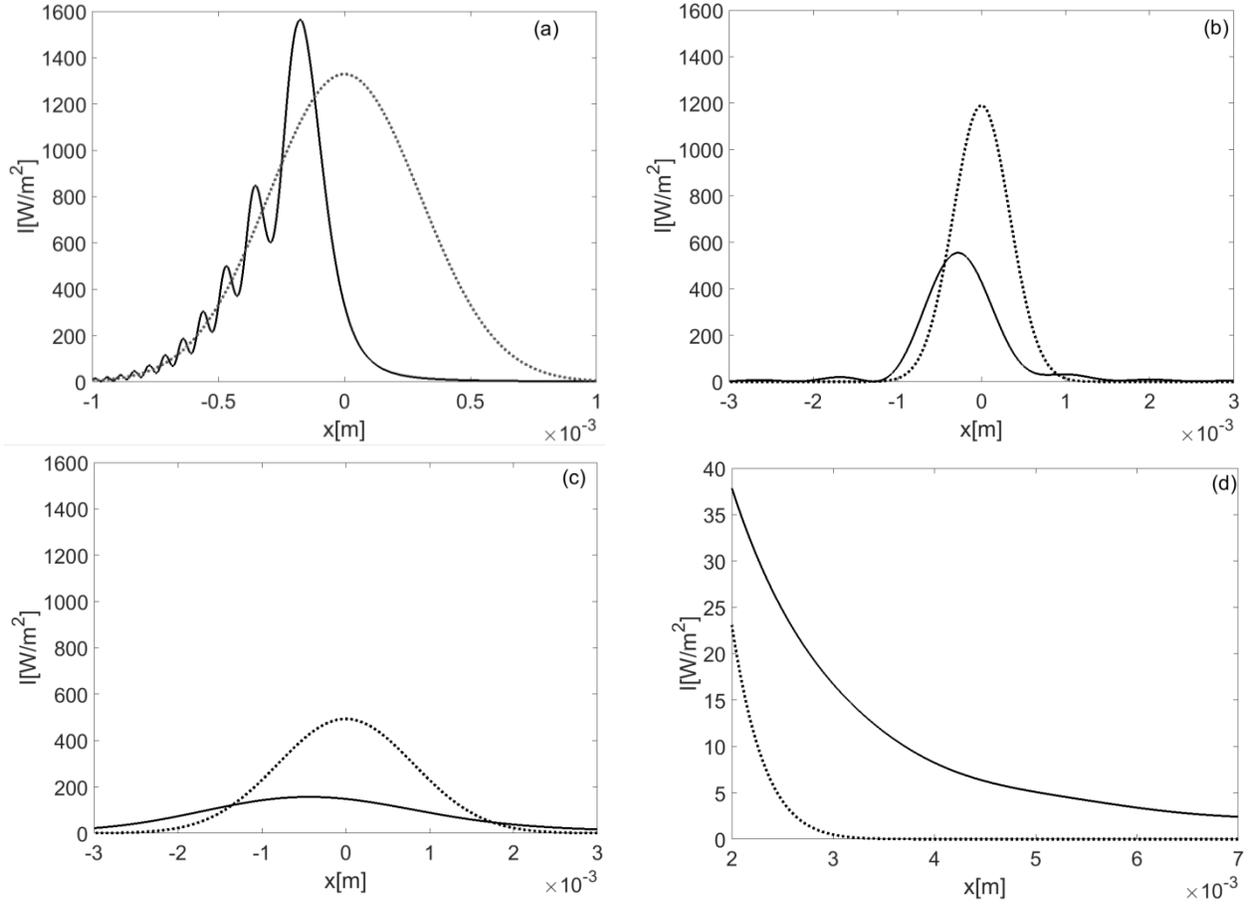

**Figure 2.** Intensity $I = U(\vec{r})c$ of the freely propagating laser beam (dotted line) and laser beam propagating behind a half plane at distances: (a) y = 0.05 m, (b) y = 0.6 m, and (c) y = 3 m. (d) Enlargement of panel (c).

The propagation of a free Gaussian beam is visualized in Fig. 3 in terms of a swarm of photon trajectories, showing the intensity profile at three different distances (from left to right, 0.2 m, 0.4 m, and 0.6 m). The trajectories are determined using the magnetic and electric fields, Eqs. (14) and (15), respectively, recast in the functional form (11), and then solving the corresponding differential equation (21). The initial conditions have been distributed following Eq. (23).

From the intensity curves shown in Fig. 3 we see that free propagating Gaussian beam spreads very slowly in the spatial range considered. The form of the photon trajectories is consistent with the spreading described by the intensity curves, with their density being maximal at the center of the beam and decreasing with the distance from the center. Indeed, the separation among trajectories is relatively small, which in compliance with the slow spreading displayed by the beam. Finally, also notice that these trajectories do not mixup, which means that the trajectories started on the region covered by one half of the initial Gaussian beam will contribute all the way down to the intensity associated with that part of the of Gaussian beam. To highlight this aspect, trajectories with initial positions distributed along positive and negative *x* have been denoted with black and gray color, respectively. Although it might seem a rather trivial issue, this well-known noncrossing behavior for Gaussian beams has an interesting counterpart when edge diffraction is considered, as seen below.

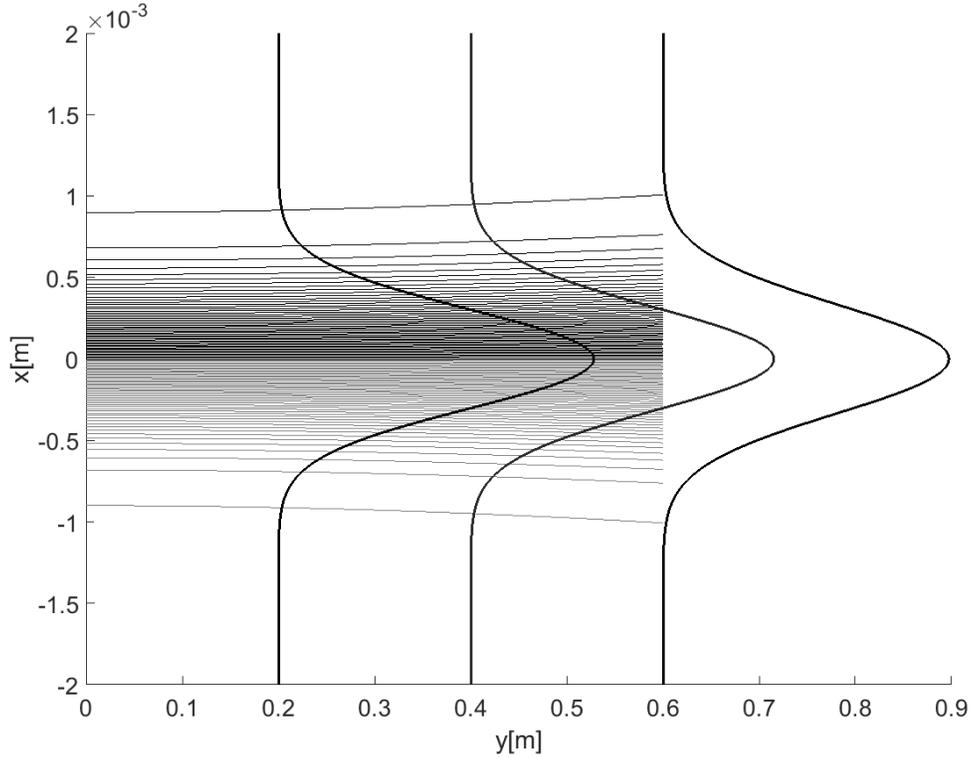

**Figure 3.** Photon trajectories for a freely propagating Gaussian beam. Intensities at three different distances are also shown (from left to right, 0.2 m, 0.4 m, and 0.6 m) to illustrate how the flow of trajectories is in compliance with the standard solution provided by the theory of diffraction. Trajectories with initial positions distributed along positive and negative x have been denoted with black and gray color, respectively, to show that they do not mixup.

The propagation of an *H*-polarized Gaussian beam obstructed by a totally absorbing half plane is shown in Fig. 4 in terms of a swarm of photon trajectories. In this case, the spreading is faster due to the diffraction undergone by the incident beam, so the trajectories are shown in two different panels corresponding to two distances from the edge [see panels (a) and (b)], which also helps us to better visualize the change in shape of the density profile (see discussion below). As before, the trajectories are determined by recasting the magnetic and electric fields, Eqs. (14) and (15), respectively, in terms of the functional form (13), and then solving the corresponding differential equation (21).

As before, we readily notice in Fig. 4 that the density of trajectory end points at a given distance from the edge is in compliance with the corresponding EME density curve displayed. Since the profile of the EME density changes, we have considered two different distances to stress such agreement, specifically y = 0.05 m and y = 0.6 m [panels (a) and (b), respectively]. Thus, the accumulation of end points is more prominent in those regions where the EME density is larger, while it is scarcer where the latter has smaller values, as seen more noticeably in panel (b). In the case of panel (a), there are two different behaviors. On the one hand, to the left of the largest maximum of the EME density, there is a monotonic decrease of the density of end points, which is somehow analogous to the behavior observed in the case of free propagation (see Fig. 3). On the other hand, to the right of such a maximum, the EME density profile displays an oscillatory behavior, which produces an also oscillatory variation of the density of end points as we move to the right, getting scarcer as we move further away from the largest maximum. This behavior is thus quite different from the one observed in the edge-free propagation.

Also, in order to emphasize the differences with respect to the case free of edge diffraction, we have considered a different color for the trajectories according to their initial distribution. Here it is we do not have a Gaussian, but half a Gaussian, as described by Eq. (20). So, taking this distribution as a whole, we have divided into two parts. Accordingly, those trajectories started between the center of the Gaussian (which coincides with the position of the edge) and some $x$ value in its decreasing part are displayed with black color, while the remaining trajectories (to the right) are with gray color. The $x$ value has been chosen in such a way that, although the initial positions are distributed according to a Gaussian, there is the same number of black and gray trajectories. Accordingly, as shown in panel (a), near the edge we observe that the contribution to the main maximum essentially arises from the black trajectories, while the gray trajectories contribute to the oscillatory part of the intensity. In other words, while the black trajectories are found to be more spatially confined, the gray trajectories are going to cover a larger region. On the other hand, as can be seen in panel (b), far from the edge (at a distance of about one order of magnitude larger than in the previous case), we observe that both groups of trajectories are nearly the same, with the black trajectories contributing to the intensity to the left of the maximum, and the gray trajectories doing it to the right, something that reminds the case of the edge-free propagation seen above. The key to understand this remarkable change of behavior is in the twist undergone by the trajectories at about twice the distance considered in panel (a), i.e., y ~ 0.1 m, which can be more clearly seen in the enlargement of panel (b) [see panel (c)]. This twist produces a spatial redistribution of the trajectories, which eventually leads to an equal distribution of trajectories on both sides of the central maximum displayed by the intensity asymptotically, i.e., the central bright spot observed in the experiment reported in Sec. 3. Furthermore, also notice that the two surrounding tails, typical of this diffraction phenomenon, are associated with the faster spreading undergone by the most marginal trajectories on both sides. Of course, the number of trajectories departing to regions far from the center of the main maximum is rather small (compared to those contributing to this maximum), which explains why the intensity in the tails so weak but spreads out over a long spatial range (perpendicularly with respect to the edge).

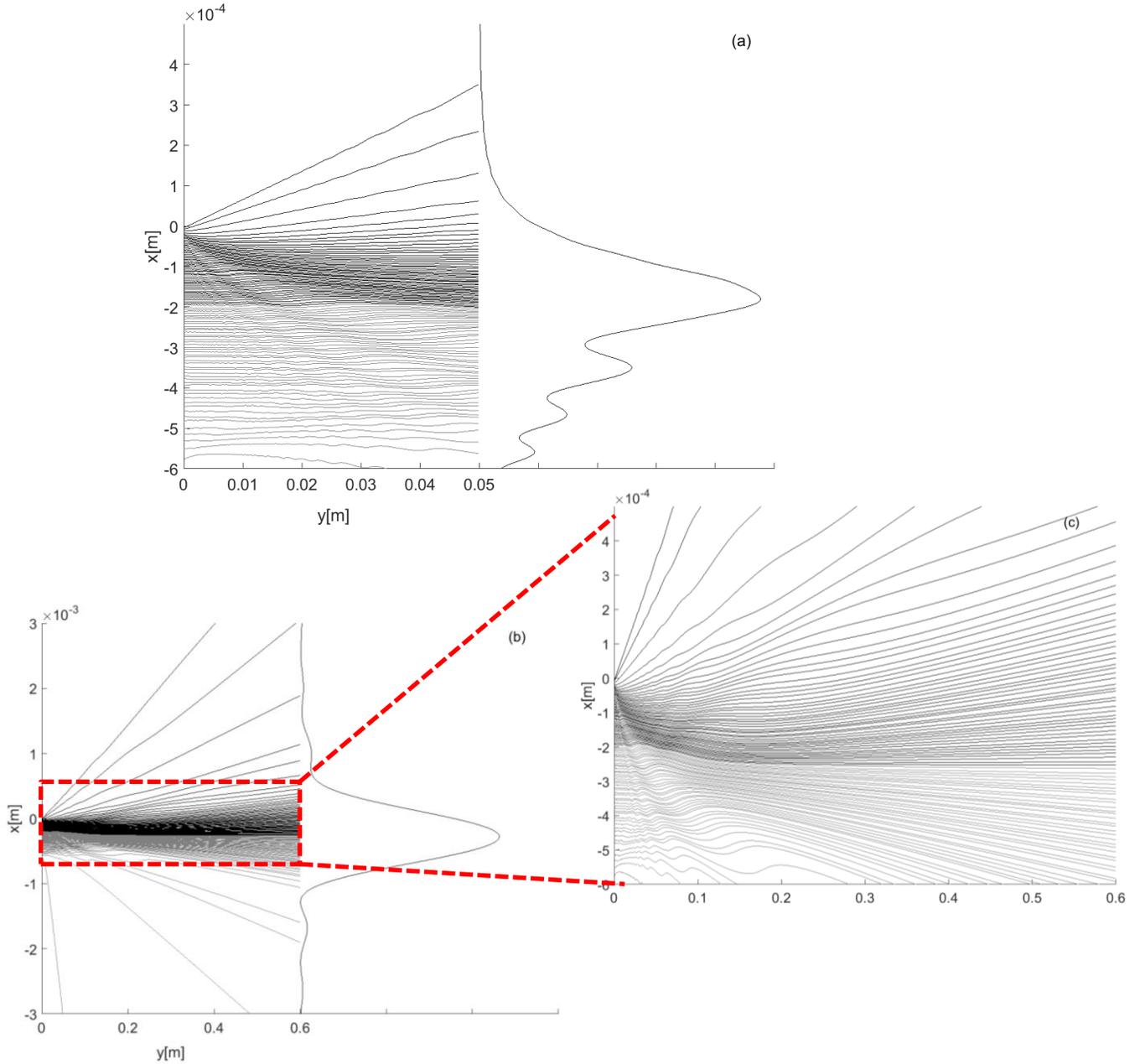

**Figure 4.** Swarms of photon trajectories illustrating the behavior of the EME flow behind a half plane illuminated by a Gaussian beam ($\sigma = 0.3$ mm) centered at the edge of the half plane. The distribution of initial coordinates for the trajectories is determined by the intensity of the incident *H*-polarized Gaussian beam. For a better visualization, trajectories associated with half of the incidence energy (near the maximum of the Gaussian) are displayed with black, while those associated with the other half of the intensity (in the decreasing part of the Gaussian) are in gray color. (a) Photon trajectories up to a distance y = 0.05 m. The intensity profile denotes the light energy density (right) evaluated from Eq. (20) at the distance y = 0.05 m. (b) Photon trajectories behind half plane up to a distance y = 0.6 m. The intensity profile denotes the light energy density (right) evaluated from Eq. (20) at y = 0.6 m. (c) Enlargement of panel (b) in order to show in more detail the central photon trajectories and the effect of the edge on them.

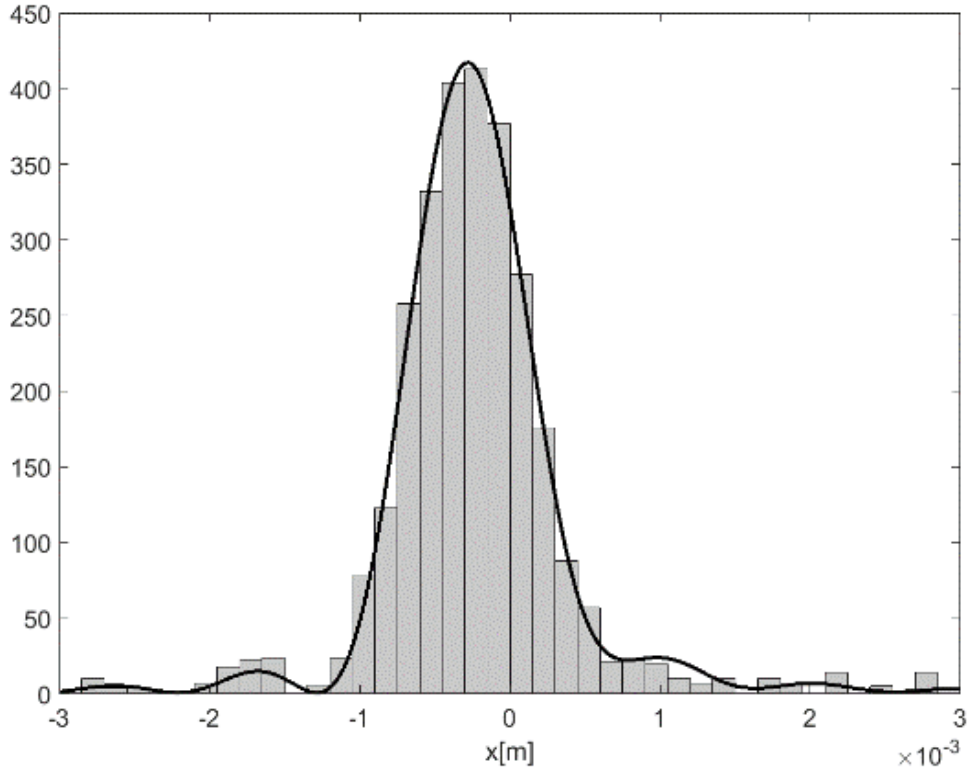

**Figure 5.** Histogram built by box-counting trajectory endpoint along the *x* axis and at a distance y = 0.6 m from the half-plane. The total number of trajectories considered to produce this statistics has been 3,000. The black solid line denotes the green light intensity profile at y = 0.6 m, obtained from Eq. (20).

In order to show the validity of the above qualitative description, we have evaluated 3,000 trajectories up to a distance y = 0.6 m from the half plane. As seen in Fig. 5, the histogram of the number of trajectory endpoints along the *x*-axis at the chosen distance shows an excellent agreement with the corresponding intensity profile. It is remarkable that the histogram of end points reproduces very well the difference in the form of the intensity profile on the left and right hand sides, even at this large distance.

## 5. Concluding remarks

When a sharp edge obstructs the passage of a laser beam, a characteristic diffraction pattern can be observed behind the edge, consisting of a bright spot and two fainter long tails leaving such a spot along the perpendicular direction to the edge. This problem has been tackled in the literature by means of the usual theory of diffraction. Here, inspired by the Bohmian formulation of quantum mechanics, we have reconsidered the problem from the viewpoint of electromagnetic flow lines, which provide a more complete interpretation of the phenomenon in terms of EM fields and flux lines, and that can be related, at the same time, with average photon paths. As has been illustrated by means of histograms, these lines or trajectories are in compliance with Maxwell's equations, showing that they are not an external element to the electromagnetic theory, but that they constitute a convenient tool to understand the evolution or propagation of electric and magnetic fields, particularly in problems of interference and diffraction.

The theoretical results presented here have been obtained on the basis of a real experiment, which has been previously performed in the laboratory making use of a relatively simple experimental setup, consisting of a green laser pointer (light source), a razor (sharp edge) and an observation screen, all of them allocated along an optical bench. As has been shown, the results were in good agreement with the experimental observations.

**Acknowledgements.** The authors acknowledge support from the Ministry of Science of Serbia under Projects No. III45016 (Milena D., M.B. and D.V.) and No. OI171028 (Milena D. and Miloš D.), and the Spanish MINECO under Project No. FIS2016-76110-P (A.S.).